\documentclass[prl,twocolumn,amsmath,amssymb,floatfix,showpacs,superscriptaddress]{revtex4}

\usepackage{graphicx}

\usepackage{bm}

\usepackage{amssymb}
\usepackage{subfigure}
\usepackage{color}
\usepackage{epsfig}
\newcommand{\up}{\uparrow}
\newcommand{\down}{\downarrow}

\def\nab{{\mbox{\boldmath{$\nabla$}}}}
\def\sig{{\mbox{\boldmath{$\sigma$}}}}

\begin{document}

\title{Effect of pair-breaking on mesoscopic
persistent currents well above the superconducting transition
temperature}

\author{H. Bary-Soroker}
\email{hamutal.soroker@weizmann.ac.il}

\affiliation{Department of Condensed Matter Physics,  Weizmann
Institute of Science, Rehovot 76100, Israel}

\author{O. Entin-Wohlman}

\affiliation{Department of Physics, Ben Gurion University, Beer
Sheva 84105, Israel}

\affiliation{Albert Einstein Minerva Center for Theoretical
Physics, Weizmann Institute of Science, Rehovot 76100, Israel}

\author{Y. Imry}

\affiliation{Department of Condensed Matter Physics,  Weizmann
Institute of Science, Rehovot 76100, Israel}

\date{\today}

\begin{abstract}
We consider the mesoscopic normal persistent current (PC) in a
very low-temperature superconductor with a bare transition
temperature $T_c^0$ much smaller than the Thouless energy $E_c$.
We show that in a rather broad range of pair-breaking strength,
$T_c^0 \lesssim \hbar/\tau_s \lesssim E_c$, the transition
temperature is renormalized to zero, but the PC is hardly
affected. This may provide an explanation for the magnitude of
the average PC's in the noble metals, as well as a way to
determine their $T_c^0$'s.
\end{abstract}

 \maketitle

\noindent{\bf Introduction.} The magnitude of the equilibrium
averaged persistent currents (PC's)~\cite{BIL,book} in normal
metals has been a long-standing puzzle. Experiments
\cite{LDDB,JMKW,DBRBM} produce a current larger by at least two
orders of magnitude than the theoretical prediction for
noninteracting electrons \cite{CGR,RvO,AGI} and seem to indicate
that the low-flux  response is diamagnetic. The average PC of a
diffusive system with interactions was calculated first in this
connection \cite{AL} in Refs.~\onlinecite{AEEPL} and
\onlinecite{AEPRL}. The Resulting PC was found to be much larger
than that of a noninteracting system, but nevertheless not large
enough to explain the experiments.

Repulsive electron-electron interactions \cite{AEPRL} result in a
paramagnetic response (at small magnetic fluxes) whose magnitude is
smaller than the experiment by about a factor of five.  This
disagreement is due to the downward renormalization of the
interaction \cite{dG,MA}. Attractive interactions~\cite{AEEPL}
result in a diamagnetic response, whose magnitude (due to the very
low superconducting transition temperature), is again smaller by a
factor of order five than the measured one. This is in spite of the
renormalization upward of the attractive interaction. Attractive
interactions, at low energies, imply (with no pair-breaking)  a
transition into a superconducting state, and the PC of such an
interacting system depends on its transition temperature. These
temperatures are very low \cite{Mota} for the noble metals used in
the PC experiments -- hence the too small predicted values for the
PC.

\begin{figure}[tbp]
\begin{center}
\includegraphics[width=8.6cm]{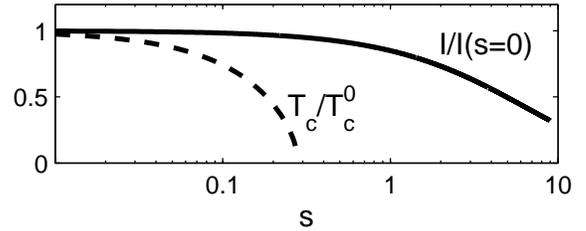}
\end{center}\vspace{-0.5cm}
\caption{The first flux harmonic [$m=1$, see Eq. \ref{MAIN}], in
units of $I(s=0)$, of the PC at $T=E_c$ (full line) and
$T_c/T_c^0$ (dashed line) as functions of the pair-breaking
strength, $s=1/(\pi T_c^0 \tau_s)$, displayed on a logarithmic
scale.} \label{fig1}
\end{figure}

Here we consider attractive interactions. We show that the presence
of a very small amount of pair breakers, e.g., magnetic impurities
(which seem to be very difficult to avoid in these metals
\cite{PGAPEB}), may change the picture profoundly. Obviously one may
consider other pair-breakers, such as a two-level systems \cite{IFS}
or simply a magnetic field \cite{SO}. In this letter we treat
specifically the case of magnetic impurities. We find that within a
significant range of the pair-breaking strength, the magnetic
impurities {\em suppress the transition temperature down to
immeasurable values, leaving concomitantly the PC almost unchanged}.
The physical reason for this remarkable observation is that the PC
is determined by the interaction on the scale of the Thouless energy
$E_c = \hbar D/L^2$ $(\sim 20 mK$ for a typical experimental
system), while the bare transition temperature, $T_c^0$, is much
smaller. (The circumference of the ring is denoted by $L$ and $D$ is
the diffusion coefficient.) This gives rise to a rather wide range
of pair-breaking strengths, presented here by the spin-scattering
time $\tau_{s}$,
\begin{align}
T_c^0 \lesssim \hbar/\tau_s \lesssim E_c
 , \label{1}
\end{align}
in which the actual  transition temperature $T_c$ will drop to
zero \cite{AG}, but the PC will be hardly affected. As a result,
it is the {\em bare} transition temperature of the system {\em
without} the magnetic impurities, $T_c^0$,  as opposed to
$T_{c}$, which dominates the expression for PC, see
Fig.~\ref{fig1}.
We concentrate here on the experimental results of
Ref.~\onlinecite{LDDB} \cite{future}. In order to explain them,
it is necessary to assume a $T_c^0$ in the $1 mK$ range for
copper. {\em Our basic assertion is that this may indeed be the
correct order of magnitude of $T_c^0$ for ideally clean copper,
but that it is knocked down to zero or to a very low value by a
minute, $\lesssim$ ppm, amount of unwanted \cite{PGAPEB}
pair-breakers.} We emphasize, however, that our result
concerning the fundamentally different sensitivities of $T_c$
and PC to pair-breaking in the range given by Eq.~(\ref{1}),
{\em remains valid regardless of the situation in specific
materials.} The Kondo screening of the spins is not considered
here. Other effects of magnetic impurities have previously been
considered in Ref.~\onlinecite{ES}.

\begin{figure}[tbp]
\begin{center}
\includegraphics[width=8.6cm]{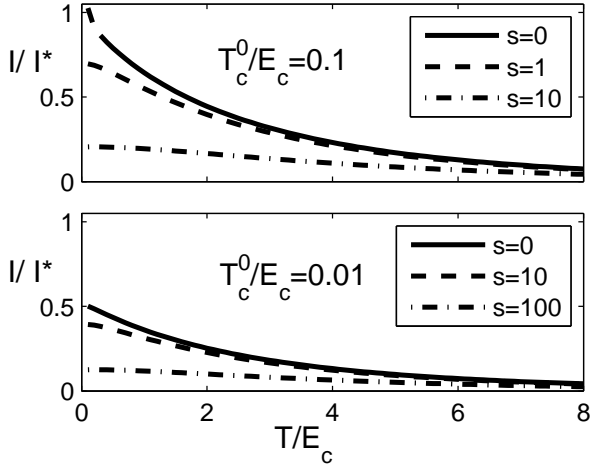}
\end{center}\vspace{-0.6cm}
\caption{The first flux harmonic of the PC in units of $I^*=- e
E_c$ as a function of the temperature, for two values of
$T_c^0/E_c$ and several values of s. Keeping, at $T\lesssim
E_c$, up to the 100 lowest values of $|\nu|$, was necessary for
convergence. Note that the $s=0$ curve in the upper panel is
valid only for $T/T_c\geq 1+Gi$, where $Gi$ is the Ginzburg
parameter ($Gi\sim0.1$ for the samples of
Ref.~\onlinecite{LDDB}).} \label{fig2}
\end{figure}
\noindent {\bf Results.} The expression we obtained for the PC
in a diffusive ring with magnetic impurities can be expressed as
a sum over the harmonics of the magnetic flux through the ring
$\phi$, in units of the flux quantum $h/e$,
\begin{align}
&I=-8eE_{c}\sum_{m=1}^{\infty} \frac{\sin(4\pi
m\phi)}{m^2}\nonumber\\
&\times\sum_\nu \int_0^\infty dx\frac{x \sin(2\pi x) \Psi
'(F(x, \nu ))}{\ln(T/T_c^0)+\Psi(F(x, \nu ))-\Psi(\frac{1}{2})} \ ,\nonumber\\
&F(x,\nu )=\frac{1}{2}+\frac{|\nu|+2/\tau_s}{4\pi T}+\frac{\pi
E_c x^2}{m^2 T}\ ,\label{MAIN}
\end{align}
(using $\hbar=1$). Here $\nu$ denotes the bosonic Matsubara
frequency \cite{COM1}, $\Psi$ and $\Psi '$ are the digamma
function and its derivative, and $T$ is the temperature. Our
expression (\ref{MAIN}) generalizes the result of
Ref.~\onlinecite{AEEPL} for the case where spin-scattering is
present: the Matsubara frequency $|\nu |$ is shifted by
$2/\tau_{s}$. However, the superconducting transition
temperature (which appears formally in the denominator of the
integrand) is {\em not} the one modified by the pair-breakers,
but retains its bare (magnetic impurities free) value.
Interestingly enough, it follows that by measuring the PC one
may determine $T_{c}^{0}$ (which would be directly measurable
only if all low-temperature pair breaking could be eliminated).

In Fig.~\ref{fig2} the PC is plotted using Eq.~(\ref{MAIN}).
At the critical pair-breaking time $1/\tau_s \simeq T_c^0$,
corresponding to $s =1/\pi \tau_s T_c^0\simeq 1/\pi$, the
transition temperature vanishes \cite{AG}, while the PC is
hardly affected. The measured PC in the copper samples of Ref.
\onlinecite{LDDB} is $I(T\lesssim E_c)\simeq -eE_c$. The curve
with $s=1$ in the upper panel, taken with $T_c^0 = 1.5 mK$ and
$E_c=15 mK$ (the value for the samples of
Ref.~\onlinecite{LDDB}) gives a PC lower by only $25\%$. A
better fit is possible by changing the parameters somewhat, but
we do not regard this as crucial at the present stage. Likewise,
we can qualitatively explain the result of
Ref.~\onlinecite{JMKW}. The high frequency results of
Ref.~\onlinecite{DBRBM} require a separate discussion
\cite{DBRBM}. The PC is reduced significantly once $1/\tau_s\geq
E_c$, or $L_s\equiv\sqrt{D\tau_s}\leq L$. For $T_c^0/E_c=0.1
\;(0.01)$, the condition for $E_c\tau_s\sim1$ is $s=10\;(100)$.

\noindent {\bf Derivation.} For completeness, we outline below
the derivation of the PC in the presence of magnetic scattering
\cite{future}. The PC, Eq.~(\ref{MAIN}), is obtained by
differentiating the free energy with respect to the flux. Our
system is described by the Hamiltonian \cite{AG}
\begin{align}
{\cal H}&=\int d{\bf r}\Bigl (\psi^{\dagger}_{\alpha}({\bf r})
\Bigl [({\cal H}_{0}+u_{1}({\bf r}))\delta_{\alpha\gamma}+u_{2}
({\bf r}){\bf S}\cdot\sig^{\alpha\gamma}\Bigr ]\psi^{}_{\gamma}({\bf r})\nonumber\\
&\ \ \ \ \ \ \ \ \ \ \ \ \ -\frac{g}{2}\psi^{\dagger}_{\alpha}
({\bf r})\psi^{\dagger}_{\gamma}({\bf r})\psi^{}_{\gamma}({\bf
r}) \psi^{}_{\alpha}({\bf r})\Bigr )\ ,\label{HAM}
\end{align}
in which the last term is the attractive interaction, of
coupling $g$. The spin components are $\alpha$ and $\gamma$,
$\sig$ is the vector of the Pauli matrices, and ${\cal
H}_{0}=(-i\nab -e {\bf A})^{2}/2m -\mu $ ($\mu $ is the chemical
potential and ${\bf A}$ is the vector potential describing the
flux through the ring). The scattering, both nonmagnetic and
magnetic, is assumed to result from $N_{i}$ point-like
impurities, such that
\begin{align}
&u_{1}({\bf r})+u_{2}({\bf r}){\bf S}\cdot\sig\nonumber\\
&\equiv \sum_{i=1}^{N_{i}}\Bigl (\delta ({\bf r}-{\bf
R}_{i})-\frac{1}{V}\Bigr ) (u_{1}+u_{2}{\bf S}^{}_{{\bf
R}_{i}}\cdot\sig )\ ,
\end{align}
where $V$ is the system volume. In averaging over the impurity
disorder one assumes that the impurity locations, ${\bf R}_{i}$,
are random, and so are  their classical spins, such that
$\langle {\bf S}^{}_{{\bf R}_{i}}\rangle =0$, and $\langle {\bf
S}^{}_{{\bf R}_{i}}\cdot{\bf S}^{}_{{\bf R}_{j}}\rangle
=\delta_{ij}S(S+1)$.

The partition function, ${\cal Z}$,  is calculated by the method
of Feynman path integrals~\cite{AS}, combined with the Grassman
algebra of many-body fermionic coherent states in terms of the
variables $\psi_\alpha({\bf r},\tau)$ ($\bar\psi_\alpha({\bf
r},\tau)$). Introducing the bosonic fields $\Delta ({\bf
r},\tau)$ via the Hubbard-Stratonovich transformation leads to
the partition function $\mathcal Z=\int D(\psi({\bf
r},\tau),\bar\psi({\bf r},\tau))D(\Delta({\bf
r},\tau),\Delta^{\ast}({\bf r},\tau)) e^{-{\cal S} }$ with
\begin{align}
{\cal S}&=\int d{\bf r}\int_{0}^{\beta} d\tau \Bigl
(\frac{|\Delta ({\bf r},\tau )|^{2}}{g}
\nonumber\\
& -\frac{1}{2}\bar{\Psi}({\bf r},\tau )G^{-1}_{{\bf r},{\bf
r};\tau ,\tau}\Psi({\bf r},\tau )\Bigr )\ ,\label{ACT1}
\end{align}
where
$\bar{\Psi}=(\bar{\psi}_{\up},\bar{\psi}_{\down},\psi_{\up},\psi_{\down})$.
The inverse Green function $G^{-1}$ (at equal positions ${\bf
r}$ and equal imaginary times $\tau$) is given by
\begin{widetext}
\begin{align}
G^{-1}_{{\bf r}={\bf r}';\tau =\tau '} =\left
[\begin{array}{cccc}
-\partial_{\tau}-h^{\phi}_{\up}&-2u_{2}S_{-} &0&\Delta\\
-2u_{2}S_{+}& -\partial_{\tau}-h^{\phi}_{\down}&-\Delta &0\\
0&-\Delta^{\ast}&-\partial_{\tau}+h^{-\phi}_{\up}& 2u_{2}S_{+}\\
\Delta^{\ast}&0&2u_{2}S_{-}&-\partial_{\tau}+h^{-\phi}_{\down}\end{array}\right
]\equiv\left [\begin{array}{cccc}\hat{G}^{-1}_{\rm p}&\ &\ &\hat{\Delta}\\
\ &\ &\ &\ \\
\ &\ &\ &\ \\
\hat{\Delta}^{\dagger}&\ &\ &\hat{G}^{-1}_{\rm
h}\end{array}\right ]\ ,\label{MAT}
\end{align}
\end{widetext}
where $h_{\alpha}^{\pm\phi}={\cal H}_{0}(\pm {\bf A})+u_{1}+{\rm
sgn}(\alpha )S_{z}u_{2}$, and $S_\pm=(S_x\pm iS_y)/2$.

The integration over the fermionic part of the action
(\ref{ACT1}) yields
\begin{align}
&{\cal  Z}=\int D(\Delta({\bf r},\tau),\Delta^{\ast}({\bf
r},\tau))\nonumber\\
&\times\exp \Bigl (\frac{1}{2}{\rm Tr}\ln (\beta G^{-1})- \int
d{\bf r }\int_{0}^{\beta} d\tau\frac{|\Delta({\bf r},\tau
)|^{2}}{g} \Bigr )\ .\label{ZwithTr}
\end{align}
In order to treat the boson fields $\Delta$, we expand ${\rm
Tr}\ln (\beta G^{-1})$ up to second order in $\Delta$. This
expansion is valid for temperatures well above the transition
temperature, and, strictly speaking, above the Ginzburg critical
region. The zeroth order is omitted as it leads to the tiny
magnitude PC of noninteracting, grand-canonical, normal metal
rings \cite{CGR}. The result in Fourier space reads (the
dependence on the magnetic flux is specified below)
\begin{widetext}
\begin{align}
{\rm Tr}\ln (\beta G^{-1})\Big |^{2\textrm{nd}}= -\sum_{{\bf
q}_1,{\bf q}_2,\nu}\sum_{{\bf k}_1,{\bf k}_2,\omega} {\rm Tr
}\left[\hat G_{\rm p}({\bf k}_1+{\bf q}_1,{\bf k}_2+{\bf
q}_2,\omega+\nu)\hat\Delta({\bf q}_2,\nu)\hat G_{\rm h}({\bf
k}_2,{\bf k}_1,-\omega)\hat\Delta^\dag({\bf q}_1,\nu)\right] \
.\label{SEC}
\end{align}
\end{widetext}
The resulting expression for the partition function may be
simplified considerably. Firstly, the terms that survive the
disorder-average in the sum of Eq.~(\ref{SEC}) are those for
which \cite{AGD} ${\bf q}_{1}={\bf q}_{2}$. Secondly, the
particle and the hole Green functions, $\hat G_{\rm p}$ and
$\hat G_{\rm h}$, [see Eq.~(\ref{MAT})] are related,
\begin{align}
\hat G_{\rm h}({\bf k},{\bf k}';\omega)=- \hat G_{\rm
p}^{t}(-{\bf k},-{\bf k}',\omega)\ ,
\end{align}
where the superscript $t$ denotes the transposed. Carrying out
the integration in Eq.~(\ref{ZwithTr}),
\begin{align}
{\cal Z}=\prod_{{\bf q},\nu}{\cal N}(0)\Bigl (
\frac{V}{g}-T\;\Pi({\bf q},\nu) \Bigr )^{-1} \ ,\label{ZwithPi}
\end{align}
where ${\cal N}(0)$ denotes the extensive density of states at
the Fermi level. The polarization is
\begin{align}
&\Pi({\bf q},\nu)= \frac{1}{2} \sum_{\omega}
\varepsilon_{\alpha\gamma} K_{\omega\alpha\gamma}({\bf q},\nu)\
\end{align}
with
\begin{align}
K_{\omega\alpha\gamma}({\bf q},\nu)&= \sum_{{\bf k}_1, {\bf
k}_2}\langle G_{\alpha\alpha '}({\bf k}_1+{\bf q},{\bf k}_2+{\bf
q},\omega+\nu)\nonumber\\
&\times \varepsilon_{\alpha '\gamma '} G_{\gamma\gamma '}(-{\bf
k}_2,-{\bf k}_1,-\omega)\rangle \ .\label{FUNK}
\end{align}
Here $\varepsilon$ is the anti-symmetric tensor,
$\varepsilon_{\alpha\alpha}=0$, and
$\varepsilon_{\up\down}=-\varepsilon_{\down \up}=1$, and $G$
denotes the particle Green function.

In Ref.~\onlinecite{AG} $K(0,0)$ was calculated using a Dyson
equation. We generalize their calculation to obtain $K({\bf
q},\nu)$ and consequently the polarization becomes \cite{future}
\begin{align}
\frac{T}{{\cal N}(0)}&\Pi({\bf q},\nu)= \Psi \Bigl
(\frac{1}{2}+\frac{
\omega_D}{2\pi T}+\frac{|\nu|+D{\bf q}^2}{4\pi T}\Bigr )\nonumber\\
&-\Psi \Bigl (\frac{1}{2}+\frac{D{\bf q}^2+|\nu|+2/\tau_s}{4\pi
T}\Bigr )\ .
\end{align}
Here $\omega_{D}$ is the cutoff frequency on the attractive
interaction, and the pair-breaking time $\tau_{s}$ is given by
\begin{align}
\frac{1}{\tau_{s}}= 2\pi {\cal N}(0)N_i S(S+1)u_2^2\ .
\end{align}

The transition temperature of the system in the {\em absence} of
pair breakers, $T_{c}^{0}$, is obtained from the ${\bf q}=0, \nu
=0$ pole of ${\cal Z}$, upon setting $1/\tau_{s}=0$,
\begin{align}
\frac{V}{g{\cal N}(0)}=\Psi\Bigl
(\frac{1}{2}+\frac{\omega_D}{2\pi T_c^0} \Bigr )- \Psi\Bigl
(\frac{1}{2}\Bigr )\ .
\end{align}
(Note that the same procedure in the {\em presence} of the pair
breaking reproduces the decrease in the transition temperature
$T_{c}$, as found in Ref.~\onlinecite{AG}.) Since $\omega_D\gg
T_c^0,T$ we may use the asymptotic expansion of the digamma
function. In this way we obtain
\begin{align}
{\cal  Z}&=\prod_{{\bf q},\nu}\Bigl [ \ln\Bigl
(\frac{T}{T_c^0}\Bigr )\nonumber\\
&+ \Psi\Bigl (\frac{1}{2}+ \frac{D{\bf q}^2+|\nu|+2/\tau_s}{4\pi
T}\Bigr )-\Psi\Bigl (\frac{1}{2}\Bigr )\Bigr ]^{-1}\
.\label{SOF}
\end{align}

Finally, the PC is given by $ I=(e/h)\;\partial T\ln{\cal
Z}/\partial \phi$. In our ring geometry, the flux enters the
longitudinal component, $q_{\parallel}$, of the vector ${\bf q}$
as
\begin{align}
 q_{\parallel}=\frac{2\pi}{L}(n+2\phi)\ ,
\end{align}
where $n$ is an integer. Only the zero transverse momentum
contributes significantly to the current. Our result
(\ref{MAIN}) is obtained upon inserting Eq.~(\ref{SOF}) into the
definition of the current and employing the Poisson summation
formula. It then follows from Eq.~(\ref{MAIN}) that values of
$\tau_s$ which are detrimental to $T_c$, may hardly affect the
PC (see Fig.~\ref{fig1}).

We conclude by further explaining the physical argument behind
our result. Very roughly, the renormalization of the
dimensionless attractive interaction $\lambda$ $(>0)$ from a
higher frequency scale $\omega_>$ to a lower one, $\omega_<$, is
given by $\lambda (\omega_<) = \frac {\lambda (\omega_>) }{1-
\lambda( \omega _>)ln(\frac{\omega_>} {\omega_<})}\;$. At
$T^0_c$ and $1/\tau_s=0$, the attractive interaction should
diverge. Using this to eliminate $\lambda (\omega_D$) ($\equiv g
N(0)/V$), we obtain that for $T^0_c \lesssim \omega <<\omega_D,\
\lambda(\omega) \backsim 1/ln (\omega /T_c^0)$, which around the
Thouless scale, is close to the value found in
Ref.~\onlinecite{AEEPL}. The pair-breaking stops the
renormalization at $1/\tau_s$, but does not significantly change
the interaction on the much larger scale of $E_c$. Our
prediction can also be tested with very small rings made of
known low $T_c$ superconductors.

We point out that the mechanism suggested by Kravtsov and
Altshuler~\cite{KA}, relating extrinsic dephasing to an enhanced
PC, is different than ours, since it relies on the
rectification of the noise. \\

{\bf Acknowledgements:} We thank  E. Altman, A. M. Finkel'stein,
L. Gunther, K. Michaeli, A. C. Mota, F. von Oppen, Y. Oreg, G.
Schwiete and A. A. Varlamov for very helpful discussions. This
work was supported by the German Federal Ministry of Education
and Research (BMBF) within the framework of the German-Israeli
project cooperation (DIP), and by the Israel Science Foundation
(ISF).



\begin{thebibliography}{999}

\bibitem{BIL}
M. B\"{u}ttiker, Y. Imry, and R. Landauer, Phys. Lett. {\bf
96A}, 365 (1983).

\bibitem{book}
Y. Imry, {\it Introduction to Mesoscopic Physics}, 2nd ed
(Oxford University Press, Oxford, 2002).

\bibitem{LDDB}
L. P. Levy, G. Dolan, J. Dunsmuir, and H. Bouchiat, Phys. Rev.
Lett. {\bf 64}, 2074 (1990).

\bibitem{JMKW}
E. M. Q. Jariwala, P. Mohanty, M. B. Ketchen, and R. A. Webb,
Phys. Rev. Lett. {\bf 86}, 1594 (2001).

\bibitem{DBRBM}
R. Deblock, R. Bel, B. Reulet, H. Bouchiat, and D. Mailly, Phys.
Rev. Lett. {\bf 89}, 206803 (2002).

\bibitem{CGR}
H. F. Cheung, E. K. Riedel, and Y. Gefen, Phys. Rev. Lett. {\bf
62}, 587 (1989).

\bibitem{RvO}
E. K. Riedel and F. von Oppen, Phys. Rev. B {\bf 47}, 15449
(1993).

\bibitem{AGI}
B. L. Altshuler, Y. Gefen, and Y. Imry, Phys. Rev. Lett. {\bf
66}, 88 (1991).

\bibitem{AL} The fluctuation correction to the orbital magnetic
response above $T_c$ was calculated first by L. G. Aslamazov and
A. I. Larkin, Sov. Phys. JETP {\bf 40}, 321 (1975).

\bibitem{AEEPL}
V. Ambegaokar and U. Eckern, Europhys. Lett. {\bf 13}, 733
(1990).

\bibitem{AEPRL}
V. Ambegaokar and U. Eckern, Phys. Rev. Lett. {\bf 65}, 381
(1990).

\bibitem{dG}
P. G. de Gennes, {\it Superconductivity of Metals and Alloys}
(Addison-Wesley Publishing Co., 1989).

\bibitem{MA}
P. Morel and P. W. Anderson, Phys. Rev. {\bf 125}, 1263 (1962).

\bibitem{Mota}
The $T_c's$ of the noble metals were estimated using varying
amounts of alloying by R. F. Hoyt and A. C. Mota, Solid State
Commun. {\bf 18}, 139 (1976). The pair-breaking strengths in
these alloys are not precisely known.

\bibitem{PGAPEB}
F. Pierre, A. B. Gougam, A. Anthore, H. Pothier, D. Esteve, and
N. O. Birge, Phys. Rev. B {\bf 68}, 085413 (2003). This work
highlighted experimentally the role of minute amounts of
magnetic impurities  in producing the extra low temperature
dephasing in the noble metal samples.

\bibitem{IFS}
Y. Imry, H. Fukuyama, and P. Schwab, Europhys. Lett. {\bf 47},
608 (1999).

\bibitem{SO}
G. Schwiete and Y. Oreg, submitted in parallel with the present
work, have considered the ``strong" Little-Parks effect
including fluctuations for rings shorter than the coherence
length. There the $T_c$ is driven to zero due to the pair
breaking effect of the flux. Like in our case, the PC can still
be large outside the superconducting regime. These results are
relevant to recent experiments in Al rings, see N. C. Koshnick,
H. Bluhm, M. E. Huber, and K. A. Moler, Science {\bf 318}, 1440
(2007).

\bibitem{AG}
A. A. Abrikosov and L. P. Gorkov, Soviet Physics JETP {\bf 12},
1243 (1961).

\bibitem{future}
More details about the derivation and the results, including
comparision with Refs.~\onlinecite{JMKW,DBRBM}, will be given
elsewhere.

\bibitem{ES}
U. Eckern and P. Schwab, J. of Low Temp. Phys. {\bf 126}, 1291
(2002).

\bibitem{COM1}
The classical limit i.e., retaining only the lowest Matsubara
frequency $\nu =0$, holds {\em only} when the temperature $T$ is
larger than $E_{c}$ (see Fig.~1 in Ref.~\onlinecite{AEEPL}).

\bibitem{AS}
A. Altland and B. Simons, {\it Condensed Matter Field Theory}
(Cambridge University Press, Cambridge, 2006).

\bibitem{AGD}
A. A. Abrikosov, L. P. Gorkov, and I. E. Dzyaloshinski, {\it
Methods of Quantum Field Theory in Statistical Physics}
(Prentice-Hall, Englewood Cliffs, NJ, 1963).

\bibitem{KA}
V. E. Kravtsov and B. L. Altshuler, Phys. Rev. Lett. {\bf 84},
3394 (2000).

\end{thebibliography}
\end{document}